\documentclass[review]{elsarticle}

\usepackage{lineno,hyperref}
\usepackage{amssymb}
\usepackage{amsmath}
\usepackage{slashed}
\modulolinenumbers[0]

\journal{Physics of the Dark Universe}

\begin{document}

\begin{frontmatter}

\title{The Standard Model and Dark Matter in a unified Einstein-Dirac system with discrete extra dimensions}


	\author[mymainaddress,mysecondaryaddress]{Nguyen Ai Viet
}
\ead{naviet@vnu.edu.vn}

\address[mymainaddress]{Information Technology Faculty, Dai Nam University, Hanoi, Vietnam}
\address[mysecondaryaddress]{VNU Information Technology Institute, Vietnam National University, Hanoi, Vietnam.}
\begin{abstract}
The Dirac-Einstein system of quark-leptons coupled to gravity is constructed generalized in the space-time extended with chiral and dark discrete extra dimensions. The Einstein's gravity, all the Standard Model's gauge vector, and Higgs fields together with a dark photon, whose mass is generated by a dark Higgs scalar, can be incorporated in this framework as components of the generalized vielbein. The Standard Model derived from the extended Einstein-Hilbert action must satisfy some conditions leading to the predictions of Weinberg angle, Higgs and top quark masses in excellent agreement with experiments. Since, the Dirac operator extended by discrete derivatives has resulted in a non-diagonal mass matrix, which mixes each quark-lepton with its Kaluza-Klein partner. On one hand, the small mixing angle set a Stueckelberg mass to the quark-leptons' Kaluza-Klein partners at least in the hundreds TeV range. On the other hand, it implies weak couplings between the Kaluza-Klein partners of quark-leptons and the Standard Model's gauge vector fields. This feature indicates that these Kaluza-Klein modes can be considered as candidates of dark matters.
\end{abstract}


\end{frontmatter}

\section{Introduction}
Our current knowledge of the elementary building blocks of Nature and their interactions is based on two independent pillars: the Standard Model and General Relativity. On the one hand, in the Standard Model, quark-leptons are fermions interacting with each other via the Yang-Mill vector bosons of the nonabelian gauge group $SU(3)_c \times SU(2)_L \times U(1)_Y$ representing the strong and electroweak interactions. The Higgs mechanism, which is triggered by a scalar doublet with a quartic self-interaction potential, gives masses to the quark-leptons and gauge vector bosons. On the other hand, in Einstein's General Relativity the gravity is described by a curved metric tensor of the Riemannian manifold of space-time ${\cal M}^4$. In principles, one can construct the Standard Model in a curved space-time to have an Einstein-Yang-Mills-Dirac system to describe all the known interactions of quark-leptons.

Although such a combined picture has incorporated all our current knowledge of the material world, we are not quite comfortable with several unanswered questions concerning it. In this paper, we will address at least the three most remarkable ones. Firstly, the strong and electroweak interactions are mediated by the Yang-Mills gauge bosons, while gravity is originated from the metric structure of space-time. In a unified framework, all the interactions should be derived from a single common principle. Secondly, the presence of Higgs doublet must be explained reasonably with a fundamental guiding principle. Thirdly, the latest cosmological observations have suggested that quark-leptons contribute only less than $5\%$ of the total matter in Nature. One might wonder how dark matter can come into our world picture.

The gravitational interaction of a fermionic matter field $\psi(x)$ in the traditional 3+1-dimensional space-time ${\cal M}^4$ is described by the Einstein-Dirac action
\begin{equation} \label{SDE}
	S_{ED}(4)= \int d^4x \sqrt{|det (g)|}~ [ i \bar \psi(x) \gamma^a e_a^\mu(x) (\partial_\mu + \omega_\mu(x)) \psi(x) + M^2_{Plank}(4) r_4], 
\end{equation}
where $x= (x^0, x^1, x^2, x^3), a= \dot 0,\dot 1,\dot 2,\dot 3; ~\mu = 0,1,2,3$. The dotted integers are used for the flat index $a$ to avoid confusion. $e^a_\mu(x)$ is the traditional vierbein representing gravity, while $r_4$ is its Ricci scalar curvature. $M_{Planck}(4) = 1/ 4 \sqrt{\pi G_N} $ is the Planck mass, where $G_N$ is the Newton's gravitational constant. The spin connection $\omega_\mu(x)$ is necessary to make the derivative covariant.   

In our previous work \cite{VDHW2017}, we have successfully derived the Einstein-Yang-Mills-Dirac systems from the generalized Einstein-Dirac action 
\begin{eqnarray} \label{SDE5}
		S_{ED}(5) &=& S_D(5) + S_{HE}(5) \nonumber \\
		S_D(5) &=& \int d^5x \sqrt{|det (G)|}~ [ \bar \Psi(\hat x)(i  \Gamma^A  E^M_A({\hat x}) (D_M + \Omega_M (\hat x)) \Psi(\hat x) \nonumber \\
		S_{HE}(5) &=&  M^2_{Planck}(5) Tr \int d^5x \sqrt{|det (G)|}  R_5,
\end{eqnarray}
where $\hat x = (x, x^5), A= a, \dot 5, M = \mu, 5$. $\Psi(\hat x)$,  $G$, $ \Gamma^A$, $E^M_A({\hat x})$, $R_6$, $ D_M$, $\Omega_M(\hat x)$ and $M_{Planck}(5)$ are respectively the quark-lepton spinors, metric, flat Dirac matrices, vielbein, Ricci scalar curvature, partial derivatives, spin connection and Planck mass in the extended space-time ${\cal M}^5$ with a discrete extra dimension. This result is encouraging because the strong and electroweak interactions' Yang-Mills gauge fields in this framework have emerged as Kaluza-Klein partners of the traditional gravity. Due to the special property of the chiral matrix $\gamma^5$, the discrete derivative has just led to a mass term of fermions. Since such a unified framework contains fewer free parameters, it leads to some phenomenological predictions, which are in good agreement with the experimental observations. In particular, the Weinberg angle is predicted as $\sin^2 \theta_W=0.23077$, suggesting that the theory's energy scale is exactly the electroweak one. Since this framework just allows to unify QCD and electroweak interaction separately, without the Higgs field, it must be extended. 

In this paper, we will go further in this direction to generalize the action in Eq.(\ref{SDE5}) in the space-time ${\cal M}^6$, which is an extension of ${\cal M}^5$ with one more discrete extra dimension to incorporate all the Yang-Mills gauge vector and Higgs fields of the Standard Model as components of the generalized vielbein. The Einstein-Dirac action of ${\cal M}^6$ will be given in the following form 
\begin{eqnarray} \label{SDE6}
	S_{DE}(6) &=& S_D(6) + S_{HE}(6) \nonumber \\
	S_D(6) &=& \int d^6x \sqrt{|det ( {\bf G})|}~ i\bar {\bf \Psi}(\tilde x) {\bf \Gamma}^E {\bf E}^P_E({\tilde x}) ({\bf D}_P + {\bf \Omega}_P(\tilde x)){\bf \Psi}(\tilde x) \nonumber \\
	S_{HE}(6) &=& M^2_{Planck}(6)~ Tr \int d^6x \sqrt{|det ({\bf G})|}~{\bf R}_6,
\end{eqnarray}		
where $\tilde x = \hat x, x^6, E= A, \dot 6, P= M, 6$. The quantities  ${\bf G}$, ${\bf \Gamma}^E$, ${\bf E}^P_E({\tilde x})$, ${\bf R}_6$, ${\bf D}_P$ and $M_{Planck}(6)$ are respectively the metric, flat Dirac matrices, vielbein, Ricci scalar curvature, partial derivatives and Planck mass in the extended space-time ${\cal M}^6$.  ${\bf \Omega}_P$ is the generalized spin connection. The extended spinor ${\bf \Psi}(\tilde x)$ now contains both quark-leptons and their Kaluza-Klein siblings. Since in the $4+1$-dimensional space-time, it is not possible to define a chiral Dirac matrix, the second discrete dimension has brought new features into this framework. It has been recently observed \cite{Viet2020b} that the discrete derivative along the new non-chiral dimension leads to the mixing of fermions and gauge vectors fields, which implies that the Kaluza-Klein partners of quark-leptons interact weakly with the gauge vector bosons like the dark matter does.

The framework for extending the space-time manifold ${\cal M}^4$ with discrete dimensions is based on the so-called Discretized Kaluza-Klein theory (DKKT), which has been developed since long time ago \cite{Viet1994, Viet1996a, Viet1996b, Viet2003, VietDu2017}. Originally, DKKT has been constructed based on the rigorous mathematical foundation of noncommutative geometry (NCG) \`a la Connes \cite{ConnesBook1994}, which is a natural extension of Riemannian geometry. However, it has been soon recognized that this approach is completely equivalent to a formal extension of the space-time ${\cal M}^5$ with a discrete extra dimension, where both fermions and vierbein are accompanied by their Kaluza-Klein partners. So, DKKT can be formulated in parallelism with the general relativity and quantum field theory step by step with transparent intuitive physical meanings without resorting to sophisticated mathematical notions. 

In the traditional geometry, ${\cal M}^5 \sim {\cal M}^4 \times Z_2$ can be treated as a 4-dimensional manifold. In the context of NCG \`a la Connes \cite{ConnesBook1994} and DKKT, the additional derivatives representing new dimensions have brought about new features making this geometric structure able to provide richer content than a mere product of space-time with a set of points. The main motivation of using discrete dimensions over the continuous ones in unified theories is that they do not imply infinite towers of massive fields, so will be more predictive. It is worth noting that the notion of discrete dimension has also been used by other authors \cite{AS2004,Deffayet2004,Deffayet2005,Kan_2003, Randall_2005,Gallicchio_2006} in various contexts by an ad hoc notion of discrete derivative.

In this paper, for convenience, we will also use the following index and coordinate conventions
\begin{eqnarray}
	\tilde x &=& (\hat x, x^6);~~\hat x = (x, x^5);~~ x=(x^0,x^1, x^2, x^3) \nonumber\\
	E,F,G,H &=&(A,\dot 6);~~ A,B,C,D= a, \dot 5;~~ a,b,c,d = {\dot 0, \dot 1, \dot 2, \dot 3} \nonumber \\
	P,Q,R,S &=&(M,6);~~M,N,L,K =(\mu,5);~~\mu,\nu,\rho, \tau=(0,1,2,3).
\end{eqnarray}

The paper is organized as follows: In Section 2, we reconstruct the generalized gauge theory, Hilbert-Einstein action, and Einstein-Dirac system in the space-time manifold ${\cal M}^5$, which are important for further extensions in Section 3. The results have been recalculated in a particular case of interest. Those who are interested in the rigorous and detailed derivations in more general cases can refer to the previous papers \cite{VDHW2017} and references therein. In particular, the Higgs and top quark masses have been revised with an improved method. In Section 3, we will define all necessary extended objects of ${\cal M}^6$ in Eq.(\ref{SDE6}). Since $S_{HE}(6)$ is also reduced to a sum of $S_{HE}(5)$ and a gauge action of ${\cal M}^5$, we can use the results of Section 2 to derive a finite content model of all interactions and Higgs field in the usual ${\cal M}^4$. Due to the second discrete so-called dark dimension, the Dirac operator in ${\cal M}^6$ results in a non-diagonal mass matrix. The diagonalizing transformation, which mixes the fermions, leads to the fermion-vielbein couplings, depending on a mixing angle. With a small value of it, the couplings of the known gauge bosons with the quark-lepton's Kaluza-Klein partners will become very small. This property is exactly what we expect from dark matter. In Section 4, we specifically focus on the various choices of realistic models. In Section 5, we summarize the results and discuss some physical implications and future directions.

\section{The space-time extended by one discrete chiral extra dimension}

The idea of using an extra dimension has been originated from Kaluza \cite{Kaluza1921} and Klein \cite{Klein1926a, Klein1926b}, when the space-time manifold ${\cal M}^4$ is extended to ${\cal M}^4 \times S^1$ with an additional compact extra dimension as a circle $S^1$ with a small radius $R$. The Fourier coefficients of all extended fields are identified as the physical modes having masses proportional to $1/R^2$. Therefore, one can argue that if $R$ is very small, these modes are not observed at the currently accessible energy. However, the infinite towers of massive modes are theoretically problematic and hardly predictive. The most beautiful result of the traditional Kaluza-Klein theory is that electromagnetism emerges as a component of the 5-dimensional metric representing gravity.

The concept of discrete extra dimension is attractive since it keeps the most important features of the traditional Kaluza-Klein theory without having to include the unwanted infinite towers of massive modes.  

\subsection{Spinor representation and Dirac operator}
It is Connes and Lott \cite{CoLo1989}, who have proposed to arrange the left- and right-handed chiral quark-leptons into the following 2-column spinor 
\begin{equation} \label{2crep}
	\Psi(\hat x) = \begin{bmatrix}
		\psi_L(x) \\
		\psi_R(x)
	\end{bmatrix}.
\end{equation} 
Essentially, the space-time is extended by one discrete extra dimension having two points to ${\cal M}^5$. Specifically, the quark-leptons with a given chirality are defined at two points of the chiral dimension. Therefore, we can conveniently represent the chiral spinors as in Eq.(\ref{2crep}).

In the Standard Model, the quark-lepton family with the index $A=1,...,N_F$ is represented in the following Connes-Lott's representation 
\begin{equation} \label{SMql}
	\Psi_A(\hat x) = \begin{bmatrix}
		\psi_{LA} (x) \\
		\psi_{RA} (x)
	\end{bmatrix},~~ 
\psi_{LA} = \begin{bmatrix}
	q^c_{L}(x) \\
	l_{L}(x)
\end{bmatrix}_A,~~
 \psi_{RA} = \begin{bmatrix}
 	u^c_{R}(x) \\
 	d^c_{R}(x) \\
 	e_{R}(x) \\
 	\nu_{R}(x)
 \end{bmatrix}_A,~~ 
\end{equation}
where c=r,y,b is the color index of quarks. $q^c_L (x)$ and $l_L(x)$ are quark and lepton doublets. So, all the operators acting on the spinors are represented by $16N_F \times 16N_F$ matrices without counting the usual Dirac spinor index $\alpha =1,2,3,4$. 




The Dirac operator acting on each quark-lepton in the spinor representation of Eq.(\ref{SMql}) is given by the following matrix \cite{VDHW2017} 
\begin{eqnarray}
	\slashed D &=& \Gamma^M D_M = (\slashed \partial \otimes {\bf 1_2} + m_1 \gamma_5 \otimes \sigma_2) \otimes {\bf 1}_{8N_F} \nonumber \\
	&=& \begin{bmatrix}
		\slashed \partial & - i m_1 \gamma_5\\
		i m_1 \gamma_5 & \slashed \partial
	\end{bmatrix} \otimes{\bf 1}_{8N_F},~~ M=\mu,5.
\end{eqnarray}
where $\sigma_a, a=1,2,3$ are the Pauli matrices, ${\bf 1}_n$ the is n-dimensional unit matrix.	

The extended Dirac matrices, which span the differential 1-forms, can be formally generalized as follows
\begin{equation}
	\Gamma^\mu = \gamma^\mu \otimes {\bf 1_2} \otimes {\bf 1}_{8N_F};~~ \Gamma^5 = \gamma^5 \otimes \sigma_2 \otimes {\bf 1}_{8N_F}.
\end{equation}

 The Dirac matrices $\Gamma^M$ satisfy the 4+1-dimensional Dirac's anti-commutation relations
\begin{equation}
	\{\Gamma^M, \Gamma^N\} = 2~ diag (-1,1,1,1,1).
\end{equation}

It is straightforward to extend the usual Dirac Lagrangian $\bar \psi(x) i \slashed \partial \psi(x)$ to
\begin{equation}
	L_{D}(5) = Tr \bar \Psi(\hat x) i \slashed D \Psi(\hat x) = \bar \psi(x) (i \slashed \partial + m_1) \psi(x),
\end{equation}
where $\psi(x) = \psi_L(x) + \psi_R(x)$. So, the chiral discrete extra dimension generates a Stueckelberg mass term in the Dirac Lagrangian. In the $4+1$-dimensional space-time ${\cal M}^5$, it might be cumbersome to explain why quark-lepton masses in the Standard Model are completely by the Higgs mechanism.

The scalar function operators in ${\cal M}^5$ are represented as follows
\begin{equation}
	F(\hat x) = \begin{bmatrix}
		f_L(x) & 0 \\
		0 & f_R(x)
	\end{bmatrix},
\end{equation} 

The action of the Dirac operator on the operator is defined as
\begin{eqnarray} \label{DF}
	\slashed DF(\hat x) &=& [\slashed D, F] = \Gamma^M (DF)_M \nonumber \\
	&=& \begin{bmatrix}
		\slashed \partial f_L(x) & im_1 \gamma^5 (f_L(x)-f_R(x)) \\
		 i m_1 \gamma_5 (f_L(x) - f_R(x))   & \slashed \partial f_R(x)
	\end{bmatrix} \otimes {\bf 1}_{8N_F}.
\end{eqnarray}
So we can read the partial derivatives as follows
\begin{equation}
	(DF)_\mu(x) = \begin{bmatrix}
		\partial_\mu f_L &0\\
		0& \partial_\mu f_R
	\end{bmatrix},~~
	(DF)_5(x) = \begin{bmatrix}
	m_1(f_R(x) -f_L(x)) &0\\
	0& m_1(f_L(x) -f_R(x))
\end{bmatrix},~~
\end{equation}
The derivative along the discrete dimension resembles the finite difference ratio $\Delta F/\Delta x^5 = m_1 \Delta F$. $m_1 = 1/\Delta x^5$ is a constant mass parameter in the unit ($\slashed h=c=1$), since $\Delta x^5$ is the distance between two points. We will treat the discrete derivative on equal footing with the continuous ones.

\subsection{Extended electroweak gauge potential and Higgs field}
 
The extended gauge field $B_\mu(x)$ is represented as a differential 1-form operator in the following $2 \times 2$ matrix representation
\begin{eqnarray} \label{5Vector}
	B(\hat x) &=& \Gamma^M B_M(\hat x) = \begin{bmatrix}
		\gamma^\mu b_{L\mu}(x) \otimes {\bf 1}_{8N_F} &  i \gamma^5 b_5^\dagger (x) \otimes {\bf Y}^\dagger \\
		- i \gamma^5 b_5(x) \otimes {\bf Y} & \gamma^\mu b_{R\mu}(x) \otimes {\bf 1}_{4N_F}
	\end{bmatrix}, \nonumber \\
    B_\mu(\hat x) &=& \begin{bmatrix}
    	b_{L\mu}(x) & 0 \\
    	0 & b_{R\mu}(x)
    \end{bmatrix}    ,~~ B_5(\hat x) =\begin{bmatrix}
    	b_5(x) & 0 \\
    	0 & b_5^\dagger(x)
    \end{bmatrix}
\end{eqnarray}	
where the gauge $b_{L,R \mu}(x)$ and Higgs $b_5(x)$ fields are also $2 \times 2$ matrices in the flavor index $i=u,d$. ${\bf Y}$ is the Higgs field's $8N_F \times 8N_F$ Yukawa coupling constant matrix to be defined below. In other words, the left- and right-handed gauge and Higgs fields are treated as Kaluza-Klein partners in the $4+1$-dimensional space-time ${\cal M}^5$. 

The field strength is generalized as follows
\begin{equation} \label{Fstrength}
	{\cal B} = \slashed D \wedge B + B \wedge B  = \Gamma^M \wedge \Gamma^N {\cal B}_{MN} (\hat x). 
\end{equation}

To order to keep the quartic potential of the Higgs field, the wedge product must be extended as follows
\begin{equation} \label{wedge}
	\Gamma^\mu \wedge \Gamma^\nu = - \Gamma^\nu \wedge \Gamma^\mu,~~\Gamma^\mu \wedge \Gamma^5 = - \Gamma^5 \wedge \Gamma^\mu,~~ \Gamma^5 \wedge \Gamma^5 = 1 
\end{equation}

From Eqs.(\ref{Fstrength}) and (\ref{wedge}) one reads
\begin{subequations}
	\begin{alignat}{4}
		{\cal B}_{\mu \nu} (\hat x) &=& {1 \over 2} (\partial_\mu B_\nu(\hat x) - \partial_\nu B_\mu(\hat x) + [B_\mu(\hat x), B_\nu(\hat x)]) \hskip 0.8cm \\
		{\cal B}_{\mu 5} (\hat x) &=&~~ {1 \over 2}  (\partial_\mu + (b_{R\mu}(x) - b_{L\mu}(x)) \sigma_3)  (B_5(\hat x) + m_1 {\bf 1}_2) \hskip 0.4cm \\
		{\cal B}_{55} (\hat x) &=& ~~(m_1 (b_5(x) + b^\dagger_5(x)) + b_5 (x)  b^\dagger_5(x))  {\bf 1}_2. \hskip 1.7cm
	\end{alignat}
\end{subequations}

The gauge Lagrangian is 
\begin{eqnarray} \label{Fkappa}
	L_g &=& - {1 \over f^2_\kappa} Tr <F^\dagger, F> \nonumber \\
	&=& - {1 \over f^2_\kappa} Tr({\cal  B}^\dagger_{\mu \nu} (\hat x) {\cal B}^{\mu \nu} (\hat x) - 2 {\cal B}^\dagger_{\mu 5} (\hat x) {\cal B}^{\mu 5}(\hat x) + {\cal B}^\dagger_{55} (\hat x) {\cal B}^{55}(\hat x)),
\end{eqnarray}
where $f_\kappa$ is a normalization factor to be fixed later.

The electroweak gauge fields for left- and right-handed quark lepton are expressed in terms the Yang-Mills $SU(2)_L$ gauge fields $W^i_\mu (x)$ and the abelian $U(1)_Y$ gauge field $W^0_\mu(x)$ as follows for each quark-lepton
\begin{eqnarray} \label{SMgauge}
	b_{L\mu}(x) &=& g W^a_\mu(x) {1 \over 2} \sigma^a \otimes {\bf 1}_{4N_F} - g' W^0_\mu(x){Y_L \over 2} \otimes {\bf 1}_{N_F} \nonumber \\
	b_{R\mu}(x) &=&  - g' W^0_\mu(x) {Y_R \over 2} \otimes {\bf 1}_{N_F},
\end{eqnarray}
where $Y_L, Y_R$ are hypercharge operators.

The scalar field $b_5(x)$ is given as follows
\begin{eqnarray} \label{SMHiggs}
	b_5(x) &=& g h_5(x) - m_1 {\bf 1}_2 \otimes {\bf Y}, ~~ h_5(x) = g {\hat H}(x) \otimes {\bf Y} \nonumber \\
	{\hat H}(x) &=&
	\begin{bmatrix}
	h_0(x)  & - h_1^*(x) \\
		h_1(x) & h^*_0(x) 
	\end{bmatrix}, 
~~ 
	{\bf Y} = \begin{bmatrix}
	Y^q. {\bf 1}_3 & 0  \\
	0 & Y^l  			
\end{bmatrix}, \nonumber \\ 
Y^q &=&  \begin{bmatrix}
	Y^u & 0  \\
	0 & Y^d  			
\end{bmatrix},~~Y^l =   \begin{bmatrix}
	Y^e & 0  \\
	0 & Y^\nu  			
\end{bmatrix}    
\end{eqnarray} 
where ${\bf Y}$ is the $8N_F \times 8N_F$ Yukawa coupling matrix of the Higgs field to quark-leptons. $Y^q$ and $Y^l$ are respectively $2N_F \times 2N_F$ Yukawa coupling matrices to quarks and leptons.

The choice of the gauge and Higgs fields in Eqs.(\ref{SMgauge}) and (\ref{SMHiggs}) ensures that the Dirac Lagrangian
\begin{equation}
	{\cal L}_f = i Tr \bar \Psi (\hat x) (\slashed D-i B(\hat x)) \Psi(\hat x)
\end{equation}
will lead to the usual Standard Model's gauge and Higgs interactions with quark-leptons.

Inserting the expressions of gauge and Higgs fields in Eqs. (\ref{SMgauge}) and (\ref{SMHiggs}) we have calculated the Lagrangian ${\cal L}_g$ as follows
\begin{eqnarray}
	&{\cal L}_g &=~ -{1 \over f^2_\kappa} ({g^2 4N_f \over 4} Tr(W^{\mu \nu}(x) W_{\mu \nu}(x)) + {5g'^2 4N_f \over 6} W^{0\mu\nu}(x) W^0_{\mu \nu}(x)\nonumber \\
	&&~~~~~~ + 2  g^2 \nabla^\mu \bar H(x) \nabla_\mu H(x) Tr({\bf Y}^\dagger {\bf Y})  - V(H(x))), \nonumber \\
	&V(H(x)) &=~~~~
	{g^4 \over 2} (\bar H(x) H(x) - {2 m_1^2 \over g^2})^2 Tr[({\bf Y}^\dagger {\bf Y})^2], \label{Higgsmass}
\end{eqnarray}
where the Higgs doublet is defined as
\begin{equation}
	H(x) = {1 \over \sqrt{2}} \begin{bmatrix}
		h_0(x) \\ 
		h_1(x)
	\end{bmatrix}
\end{equation}

In order to keep correct factors of the kinetic terms for gauge and Higgs fields, we must have the following conditions
\begin{eqnarray}
	& f^2_\kappa &= ~~~4 N_F g^2,~~ g' = g \sqrt{3 \over 10},  \label{Weinberg}\\
	& Tr({\bf Y}^\dagger {\bf  Y}) &=~~~ N_F. \label{Top}
\end{eqnarray}

Eq.(\ref{Weinberg}) recovers  the Weinberg angle's prediction $\sin^2 \theta_W = { (g'/g)^2\over 1 + (g'/g)^2 } = 0.23077$ with the accuracy of 99.9\% in \cite{VDHW2017}. It is worth to compare with the result $\sin^2 \theta_W$ of a recent model based on NCG \cite{Besnard} 

This framework also leads to a prediction of the top quark mass. In fact Eq(\ref{Top}) implies
\begin{equation}
	g^2 v^2 Tr({\bf Y}^\dagger {\bf  Y}) =  \sum m^2_i \sim  3 m^2_t= N_Fv^2 g^2 = 4 N_F m^2_W,
\end{equation}
where the mass index i runs over all the quark-lepton types.
 
With $N_F=3,~m_W = 80.4~GeV$ we have the top quark prediction $m_t = 2m_W \sim 161~GeV$, while the experimental value is $173~ GeV$. A restriction on the Higgs mass can be read from Eq.(\ref{Higgsmass}) as follows
\begin{eqnarray}
	m^2_H &=& f^{-2}_\kappa g^4 v^2 Tr[({\bf Y}^\dagger {\bf Y})^2] \leq {1 \over 4} N_F g^2 v^2 = 3 m^2_W, \nonumber \\
	m_H &\leq& \sqrt{3} m_W = 139~GeV,
\end{eqnarray}
which in accordance with the experimental value $125~ GeV$. In
Section 4, we will see that the metrics of extra dimensions will improve the top quark mass's prediction. In any case, the above predictions suggest that the energy scale of this model is not too far from the electroweak one. 

In Eq.(\ref{SMgauge}), if can include the gluon fields $g_S C^a_\lambda(x) {\lambda^a \over 2}, a=1,...,8,$
as follows
\begin{eqnarray} \label{StrongGauge}
		&b_{L\mu}(x)=& {\bf 1}_{N_F} \otimes \Big [\begin{bmatrix}
		g_S C^a_\mu(x) {\lambda^a \over 2} & 0\\
		0  &  0
	\end{bmatrix} \otimes {\bf 1}_2 +  g W^i_\mu(x) {\sigma^i \over 2} \otimes {\bf 1}_4  - g' W^0_\mu(x) {Y_L \over 2}\Big ],~~~~ \nonumber \\
	&b_{R\mu}(x)=& {\bf 1}_{N_F} \otimes \Big [ \begin{bmatrix}
		g_S C^a_\mu(x) {\lambda^a \over 2} & 0\\
		0  &  0
	\end{bmatrix} \otimes {\bf 1}_2 -   g' W^0_\mu(x) {Y_R \over 2} \Big ],
\end{eqnarray}

With the above arguments, we imply the relation $g_S = 2g/3$, which cannot be valid at the currently accessible energy scale. 

\subsection{The extended vielbein}
If the chiral extra dimension exists, we must discuss an extension of gravity. Since in the usual space-time manifold ${M}^4$ gravity can be described elegantly in Cartan's vierbein formalism, it is convenient to follow this procedure step by step to construct the extended Einstein-Cartan theory.

First of all, in the curved space-time, the curvi-linear Dirac matrices $\Gamma^M$ are related to the flat ones $\Gamma^A$ by the generalized vielbein
\begin{equation}
	\Gamma^M = \Gamma^A E^M_A (\hat x),~~ \Gamma^A = \Gamma^M E^A_M(\hat x),
\end{equation} 
where $E^M_A(\hat x)$ is the inverse matrix of the vielbein $E^M_A(\hat x)$

The most general form of the hermitian vielbein is \cite{VietDu2017, VDHW2017}
\begin{eqnarray} \label{5Vielbein}
	E^a_\mu(\hat x) &=& \begin{bmatrix}
		e^a_{L\mu(x)} & 0 \\
		0 & e^a_{R\mu}(x)  
	\end{bmatrix}~~~~~~,~~ E^a_5 =0 \nonumber \\
	E^{\dot 5}_\mu(\hat x) &=& \phi(x) \begin{bmatrix}
		a_{L \mu} (x) & 0 \\
		0 & a_{R \mu}(x)
	\end{bmatrix} ,~~ E^{\dot 5}_5(\hat x) = \phi(x),
\end{eqnarray}
where $e^a_{i\mu} (x), i=L,R$ are two vierbeins. $a_{i\mu} (x)$ are two possibly non-abelian vector fields, $\phi(x)$ is a Brans-Dicke scalar. The pair of vierbeins leads to bigravity, where one gravity will be massive, the other one remains massless \cite{Viet1996b, VDHW2017}. 

For our purpose of finding a minimal realistic model, in this paper, we consider only one massless gravity with the following particular vielbein
\begin{eqnarray} \label{5Vielbein}
	E^a_\mu(\hat x) &=& e^a_{\mu}(x) {\bf 1}_2,~~ E^a_5 (\hat x) =0, \nonumber \\
	E^{\dot 5}_\mu(\hat x) &=& \lambda^2 \begin{bmatrix}
		a_{L \mu} (x) & 0 \\
		0 & a_{R \mu}(x)
	\end{bmatrix} ,~~ E^{\dot 5}_5 (\hat x)= \lambda^2,
\end{eqnarray}
where the Brans-Dicke scalar field is frozen to a positive constant $\lambda^2$.

The inverse vielbein is 
\begin{eqnarray} \label{5InvVielbein}
	E_a^\mu(\hat x) &=& e_a^{\mu}(x) {\bf 1}_2,~~ E^\mu_{\dot 5} (\hat x) =0, \nonumber \\
	E^5_a(\hat x) &=& e^\mu_a(x)  \begin{bmatrix}
		a_{L \mu} (x) & 0 \\
		0 & a_{R \mu}(x)
	\end{bmatrix} ,~~ E_{\dot 5}^5 (\hat x)= \lambda^{-2} {\bf 1}_2,
\end{eqnarray}

The generalized metric tensor can be calculated in terms of the vielbein as in the usual case
\begin{equation}
	G_{MN}( \hat x) = E^A_M(\hat x) \eta_{AB} E^B_N(\hat x),~~ \eta_{AB} = diag(-1, 1,1,1,1). 
\end{equation}

We can calculate the components of the metric as follows
\begin{eqnarray} \label{5Metric}
	G^{55}(\hat x) &=& g^{\mu \nu}(x)
	\begin{bmatrix}
		a_{L\mu}(x) a_{L\nu}(x) & 0 \\
		0 & a_{L\mu}(x) a_{L\nu}(x)
	\end{bmatrix} + \lambda^{-4} {\bf 1}_2, \nonumber \\
	G^{5\mu}(\hat x) &=& G^{\mu 5} = g^{\mu \nu}(x)
\begin{bmatrix}
	a_{L\nu}(x) & 0 \\
	0 & a_{R\nu}(x)
\end{bmatrix},~~ G^{\mu \nu} = g^{\mu \nu}(x) {\bf 1}_2 
\end{eqnarray}

The usual Cartan structure equations are generalized as follows \cite{VDHW2017}
\begin{equation} \label{CSE5}
	T^A = D\Gamma^A + \Gamma^B \wedge \Omega^A_{~B} ,~~
	R^{AB} = D\Omega^{AB} + \Omega^A_{~C} \wedge \Omega^{CB}, 
\end{equation} 
where $\Omega^{AB}$, $T^A$ and $R^{AB}$ are the generalized Levi-Civita, torsion and Ricci curvature differential forms.

Using  the minimal set of constraints \cite{Viet2014} to replace the torsion free condition, we can use the first structure equation to determine the connections in terms of the vielbein coefficients \cite{VDHW2017}. Then one can also use the second structure equation in Eq.(\ref{CSE5}) to calculate the Ricci curvature coefficients $R_{ABCD}$. Finally, the generalized scalar Ricci can be calculated utilizing the following formula
\begin{equation}
	R_5 = \eta^{AC} R_{ABCD} \eta^{BD} 
\end{equation}


With the vielbein given in Eq.(\ref{5Vielbein}), it is straightforward to reduce the generalized Hilbert-Einstein action to the usual $3+1$-d one together with a gauge term  \cite{VDHW2017, VietDu2017} 
\begin{eqnarray} 
	S_{HE}(5) &=& M^2_{Planck}(5) Tr \int d^5x \sqrt{|det(G)} | R_5 \nonumber \\
	&=& 8 N_F M^2_{Planck}(5) \lambda^2  \int d^4x \sqrt{|det(g)|} tr R_5,  \nonumber \\
	&=& S_{HE}(4) + S_g(4) \label{5SHE}\\
	S_{HE}(4) &=& M^2_{Planck}(4) \int d^4x \sqrt{|det(g)} r_4 \\
	S_g(4)&=& M^2_{Planck}(4) \int d^4x \sqrt{|det(g)} [- {1 \over 16} g^{\mu \rho}(x) g^{\nu \tau} (x){\cal F}_{\mu \nu}(x) {\cal F}_{\rho \tau} (x)]\nonumber \\
	&=& \int d^4x \sqrt{|det(g)} {\cal L}_g(4) \label{5GaugeLag}
\end{eqnarray}
where $M_{Planck}(4)= 2 \sqrt{2N_F} \lambda M_{Planck}(5)$, $M_{Planck}(5)$, $g$ and $G$ are respectively the Plank masses and metric tensors in the $3+1$- and $4+1$-dimensional space-times. 


In the gauge action $S_g(4)$, the gauge "field strength" ${\cal F}_{\mu \nu}(x)$ is given as follows
\begin{eqnarray} \label{CALF}
	{\cal F}_{\mu \nu}(x) &=& f_{L \mu \nu}(x) + f_{R \mu \nu} (x) + m ([a_{R\nu}(x), a_{L\mu}(x)] +[a_{L\nu}(x), a_{R\mu}(x)] ),\nonumber \\
	f_{I\mu \nu} (x) & =& \partial_\mu a_{I\nu}(x) - \partial_\nu a_{I \mu}(x) + m [a_{I \mu}, a_{I\nu}], I=L,R,
\end{eqnarray}
which differs from the usual field strength tensor by the last two commutator terms. So, the Lagrangian ${\cal L}_g$ is not automatically gauge invariant. 

In \cite{VDHW2017, VietDu2017} we consider two typical cases, where ${\cal L}_g$ becomes gauge invariant. In the case of electroweak interaction, if $a_{R\mu}(x)$ is an abelian gauge vector, the commutators vanish. In the case of strong interaction $a_{L\mu}(x) = \alpha a_{R\mu}(x) = \alpha g_S C^a_\mu(x) {\lambda^a \over 2}$ are non-abelian $SU(3)_c$ gauge vectors, where $\lambda^a, a=1,...8$ are the Gell Mann matrices. It has been shown \cite{VDHW2017} that the choice $\alpha=3$ leads to a consistent gauge invariant Lagrangian.  

Since vielbein coefficients are dimensionless, in Eq.(\ref{5GaugeLag}) we have introduced the physical vector fields $a_{I\mu}(x)$ by the field redefinition
\begin{equation} \label{RescaleV5}
	a_{I\mu}(x) \rightarrow {2 \over f_\kappa M_{Planck}(4)} a_{I\mu}(x),  
\end{equation} 
which lead to the usual form
\begin{equation}
	{\cal L}_g = -{1 \over 4 f_\kappa^2} g^{\mu \rho}(x) g^{\nu \tau} (x){\cal F}_{\mu \nu}(x) {\cal F}_{\rho \tau} (x),
\end{equation}
where $f_\kappa$ is a dimensionless normalization factor. 

The results of Eq.(\ref{5GaugeLag}) is remarkable since it implies that the nonabelian gauge interactions can be incorporated in the vielbein of the extended space-time ${\cal M}^5$. Thus, DKKT's framework has provided an explanation of the strong and electroweak interactions' gauge symmetry structures. In order to include the Higgs field into the framework as a component of the extended gravity, one must make the further step to the space-time ${\cal M}^6$ in Section 3.  

\subsection{The Einstein-Dirac system in ${\cal M}^5$}

The Einstein-Dirac action in Eq.(\ref{SDE5}) can be split into two parts  
\begin{equation}
	S_{ED}(5) = S_D(5) + S_{HE}(5),
\end{equation}
where
\begin{equation}
	 S_D(5)=Tr[\int d^4x \sqrt{|det (g)|} \lambda^2 \bar \Psi(\hat x) i  \Gamma^A  E^M_A({\hat x}) (D_M + \Omega_M (\hat x)) \Psi(\hat x)].
\end{equation}
It is straightforward to derive $S_D(5)$ for the specific vielbein in Eq.(\ref{5Vielbein}) from the detailed calculations for the general case given in \cite{VDHW2017}.
\begin{eqnarray}\label{SD5}
		S_D(5)&=& \int d^4x \sqrt{-det(g)}~( {\cal L}_f + {\cal L}_{\sigma} + {\cal L}_{f-g}) \nonumber\\
		{\cal L}_f&=&  {\bar \psi}(x) (i\gamma^a e^\mu_a(x) \partial_\mu   + m_1 \lambda^{-2} - {1 \over 8}\gamma^c \omega_{abc}(x) [\gamma^a, \gamma^b])\psi (x) + h.c. \nonumber \\
		{\cal L}_{f-g}&=& {2m_1 \over f_\kappa M_{Planck}(4)} \sum_{I=L,R}  {\bar \psi}_I(x) \gamma^a e^\mu_a(x)  a_{I\mu}(x) \psi_I(x) + h.c. \nonumber \\
		{\cal L}_\sigma&=& {1 \over 8 f_\kappa M_{Planck}(4)}  \bar \psi (f_{L\mu \nu}(x)+ f_{R\mu \nu}(x))\sigma^{\mu \nu} \psi(x) + h.c.,
\end{eqnarray}
where $ \sigma^{\mu \nu} = i [\gamma^\mu, \gamma^\nu].$

Let us discuss the physical meaning of the above Lagrangians. $S_\psi$ describes the chiral spinors
in the curved 4D space-time with a mass of $m_1 \lambda^{-2}$.  $S_\sigma$ is a gauge-invariant giving a very small contribution to the magnetic momenta of the chiral spinors due to the large value of the Planck mass. $S_{f-g}$ would contain the interactions between chiral spinors and gauge fields in the curved space-time ${\cal M}^4$ if $m_1 \sim f_\kappa M_{Planck}(4)/2$. So in the $4+1$-dimensional space-time ${\cal M}^5$ the Standard Model's vector gauge bosons can be introduced in two different ways. On one hand, in the $4+1$-dimensional gauge theory, the mass parameter $m_1$ is at the electroweak energy scale. The Higgs field emerges as a component of the extended gauge field.  On the other hand, in the $4+1$-dimensional Einstein's theory, $m_1$ is at the Planck scale. In this case, the Higgs field and the electroweak energy scale must be introduced independently without a guiding principle. 

These two approaches will be combined in Section 3, where the Standard Model's gauge and Higgs fields are identified as the vielbein components in the extended space-time ${\cal M}^6$. 

\section{The extended space-time with chiral and dark extra dimensions}

In this section we will reduce the formally defined Einstein-Dirac action of Eq.(\ref{SDE6}) into $4+1$-dimensional actions, then the obtained expressions to the usual $3+1$-dimensional ones utilizing the results of Section 2. 

\subsection{Mixing of a quark-lepton with its Kaluza-Klein partner in the flat ${\cal M}^6$}

In the flat space-time ${\cal M}^6$, the extended spinor of each quark-lepton type f, where $f$ specifies the color, flavor, and generation, as in Eq.(\ref{SMql}) is given as follows
\begin{equation}
	{\bf \Psi}^f({\tilde x}) = \begin{bmatrix}
		\Psi^f_w(\hat x) \\
		\Psi^f_v(\hat x)
	\end{bmatrix} = \begin{bmatrix}
	\psi^f_{wL}(x) \\
	\psi^f_{wR}(x) \\
	\psi^f_{vL}(x) \\
	\psi^f_{vR}(x) \\
\end{bmatrix} ,
\end{equation}\
where $\psi^f_I(x), I= (w,L),(w,R), (v,L), (v,R)$ are chiral spinors of ${\cal M}^4$. Therefore, in the space-time ${\cal M}^6$, a given chiral quark-lepton of type f is extended to a Kaluza-Klein pair of w- and v-types. As we will see, each quark-lepton and its Kaluza-Klein sibling will be linear combinations of these two fermion types. 

In the space-time ${\cal M}^6$, the Dirac matrices are extended as follows
\begin{eqnarray} \label{FlatGamma}
	&{\bf \Gamma}^\mu &= \gamma^\mu \otimes {\bf 1}_2 \otimes \sigma_3,~~{\bf \Gamma}^5 = \gamma^5 \otimes \sigma_2 \otimes \sigma_3,~~{\bf \Gamma}^{\dot 6} = {\bf 1}_4 \otimes \sigma_2 \otimes \sigma_1, \nonumber \\
	 &\{{\bf \Gamma}^P, {\bf \Gamma}^Q\} &= 2 {\bf G}^{PQ} = 2~ diag(-1,1,1,1,1,1);~~ P,Q = \mu,5,6, 
\end{eqnarray}
where the first matrix acts on the usual spinor index, while the second and third ones do respectively on the chiral and dark discrete dimensions. We have omitted the unit matrix ${\bf 1}_{8 NF}$ in the above expressions since the Dirac matrices do not mix the quark-lepton types. 

The partial derivatives of the space-time ${\cal M}^6$ are formally given as follows
\begin{eqnarray} \label{PDer}
	{\bf D}_\mu & =&\partial_\mu. {\bf 1}_4 \otimes  {\bf 1}_2 \otimes \sigma_3,~~ 		
	{\bf D}_6=  m {\bf 1}_4 \otimes {\bf 1}_2 \otimes \sigma_3 \nonumber \\
	{\bf D}_5& =& - {\bf 1}_4 \otimes {\bf 1}_2 \otimes \begin{bmatrix}
		m_1  & 0 \\
		0 & - m_2	
	\end{bmatrix},
\end{eqnarray}
where $m_1, m_2, m$ are mass parameters characterizing the sizes of chiral and dark dimensions.
 
Since the generalized Dirac operator is given as follows
\begin{eqnarray}
	\slashed {\bf D} &=&{\bf \Gamma}^P {\bf D}_P = \begin{bmatrix}
		\slashed D_w & i m \sigma_2\\
		-i m \sigma_2    &  \slashed D_v
	\end{bmatrix} \nonumber \\
&=& 
\begin{bmatrix}
	\slashed \partial & i m_1 \gamma^5 & 0 & -m \\
	-i m_1 \gamma^5 & \slashed \partial& m & 0 \\
	0 & m & \slashed \partial &  i m_2 \gamma^5 \\ 
	-m & 0 & - i m_2 \gamma^5 & \slashed \partial
\end{bmatrix},
\end{eqnarray} 
the generalized Dirac Lagrangian is extended to
\begin{equation} \label{DiracL}
{\cal L}_D =  Tr \bar {\bf \Psi}(\tilde x) i \slashed {\bf D} {\bf  \Psi}(\tilde x) = \sum_{I} \bar \psi_I(x) i \slashed \partial \psi_I + \sum_{I,J} \bar \psi_I(x) {\cal M}^{IJ} \psi_J(x),
\end{equation}
where $I,J=(w,L), (w,R), (v,L), (v,R)$
\begin{equation}\label{M0M}
{\cal M} = \sigma_2 \otimes M,~~ M= \begin{bmatrix}
m_1 & -im \\
im & m_2
\end{bmatrix}.
\end{equation}

The partial derivative along the discrete dark dimension leads to non-diagonal elements of the mass matrix ${\cal M}$. As its consequence, the spinors $\Psi^i(\hat x), i=v, w$ are not mass eigenstates. Since this matrix is hermitian, it can be diagonalized by the following unitary transformation \cite{Viet2020b} 
\begin{equation} \label{Utrans}
	U = \begin{bmatrix}
		\cos \theta & i \sin \theta \\
		-i \sin\theta & -\cos \theta
	\end{bmatrix}, ~~
	{M}' = U  M U^{\dagger} = \sigma_2 \otimes \begin{bmatrix}
		\mu & 0 \\
		0 & \mu_{D}
	\end{bmatrix}, 
\end{equation}
where  $\theta$ is the mixing angle between the quark-leptons' Kaluza-Klein siblings. $\mu$ and $\mu_D$ are the real mass eigenvalues corresponding to the mass eigenstates defined as follows
\begin{equation} \label{PsiU}
 	\Psi^f_U(\tilde x) = U \Psi^f(\tilde x) = \begin{bmatrix}
 		\Psi^f(\hat x) \\
 		\Psi^f_D(\hat x)
 	\end{bmatrix},
\end{equation}
where $\Psi^f(\hat x)$ will be identified with the quark-lepton of type f represented in Eq.(\ref{2crep}), while $\Psi^f_D(\hat x)$ with its dark partner. From now on the U index will be omitted, hopefully without confusion.

Transforming ${M}'$ back to ${M}$ we have 
\begin{equation} \label{Mf}
	M =
	\begin{bmatrix}
		\mu + (\mu_D - \mu) \sin^2 \theta & {i\over 2} (\mu_D-\mu) \sin 2 \theta  \\
		-{i\over 2} (\mu_D-\mu) \sin 2 \theta & \mu + (\mu_D-\mu) \cos^2 \theta 
	\end{bmatrix}. 
\end{equation}

Comparing Eqs.(\ref{M0M}) and (\ref{Mf}) we obtain the mass splitting formula in \cite{Viet2020b}.
\begin{equation}
	\mu_D-\mu = {2 m \over \sin 2\theta}.  
\end{equation}

So, we can express the mass eigenvalues in terms of $m_1, m$ and $\theta$ as follows
\begin{equation}
	\mu= m_1 - m {\cos^2 2 \theta \over \sin 2 \theta},~~ \mu_D = m_1 + m { 1 + \sin^2 2 \theta \over \sin 2 \theta}.
\end{equation}

The mass parameter $m_2$ of the chiral dimension in the v-type sector has been replaced by the mixing angle $\theta$. The Stueckelberg masses $\mu, \mu_D$ do not depend on the quark-lepton type. The Higgs mechanism will add different additional masses to quark-leptons and their Kaluza-Klein partners. In this framework, we can assume the quark-lepton's Stueckelberg mass $\mu$ to be 0. We will give a more detailed analysis later when the gravity is also included. 

\subsection{The generalized Hilbert-Einstein action in ${\cal M}^6$}
In the curved space-time ${\cal M}^6$, the curvilinear Dirac matrices ${\bf \Gamma}^P$ satisfy the curved commutation relation
\begin{equation}
	\{{\bf \Gamma}^P,{\bf \Gamma}^Q \} = 2 {\bf G}^{PQ} (\tilde x),~~ P, Q= \mu, 5, 6.
\end{equation}
where the metric ${\bf G}^{PQ}(\tilde x)$ is space-time dependent and in general non-orthonormal.

At each space-time point $\tilde x \in {\cal M}^6$, there exist a local vielbein matrix, which transforms the curvilinear Dirac matrices ${\bf \Gamma}^P$ into the flat ones
\begin{equation}
	{\bf \Gamma}^E = {\bf \Gamma}^P {\bf E}^E_P(\tilde x), E= a, \dot 5, \dot 6, P= \mu, 5, 6,
\end{equation} 
where the flat Dirac matrices satisfy the following commutation relations
\begin{equation}
	\{{\bf \Gamma}^E, {\bf \Gamma}^F \} = 2 \eta^{EF} = 2~ diag(-1,1,1,1,1,1).
\end{equation} 

The most general vielbein of the space-time ${\cal M}^6$ is given as
\begin{eqnarray} \label{6Vielbein}
	{\bf E}^A_M (\tilde x) &=& \begin{bmatrix}
			E^{wA}_M (\hat x) & 0 \cr
		0 & 	E^{vA}_M (\hat x)
	\end{bmatrix}
	 ~,~ {\bf E}^A_6 = 0\nonumber \\
	{\bf E}^{\dot 6}_M (\tilde x) &=& \lambda^2_D \begin{bmatrix}
		B_M (\hat x) & 0 \cr
		0 & X_M(\hat x)
	\end{bmatrix} ~,~
	{\bf E}^{\dot{6}}_6 (\tilde x) = \begin{bmatrix}
		\lambda^2_D & 0 \cr
		0 & \lambda^2_D
	\end{bmatrix},
\end{eqnarray}
where $E^{IA}_M(\hat x), I=w,v$ are vielbein for both copies of the space-time ${\cal M}^5$. $B_M(\hat x)$ and $X_M(\hat x)$ are a pair of ${\cal M}^5$ vectors, where we will incorporate the gauge vector fields in a realistic model, while $\lambda_D$ is the metric parameter of the dark dimension.

The generalized metric tensor can be expressed in terms of the vielbein coefficients as follows
\begin{eqnarray}
	{\bf G}^{PQ}(\tilde x) &=& {\bf E}^P_E(\tilde x) \eta^{EF} {\bf E}^Q_F(\tilde x), \nonumber \\
	{\bf G}_{PQ}(\tilde x) &=& {\bf E}_P^E( \tilde x) \eta_{EF} {\bf E}_Q^F(\tilde x),~~ E,F = a, \dot 5, \dot 6;~~ P,Q= \mu, 5, 6.
\end{eqnarray}
where $\eta^{EF} = diag(-1,1,1,1,1,1)$. 

The wedge product of the differential forms is defined via the Dirac matrices's one as an extension of Eq.(\ref{wedge}) as follows
\begin{eqnarray}
	&&	{\bf \Gamma}^a \wedge {\bf \Gamma}^E = - {\bf \Gamma}^E \wedge {\bf \Gamma}^a, ~~	{\bf \Gamma}^5 \wedge {\bf \Gamma}^6 = - {\bf \Gamma}^6 \wedge {\bf \Gamma}^5 \nonumber \\
	&&	{\bf \Gamma}^5 \wedge {\bf \Gamma}^5 = {\bf \Gamma}^6 \wedge {\bf \Gamma}^6 = {\bf 1}_2 \otimes {\bf 1}_2
\end{eqnarray}

Similarly to Eq.(\ref{CSE5}), we have the generalized Cartan structure equation
\begin{equation} \label{CSE6}
	{\bf T}^E = {\bf D}{\bf \Gamma}^E + {\bf \Gamma}^F \wedge {\bf \Omega}^E_{~F} ,~~
	{\bf R}^{EF} = {\bf D} {\bf \Omega}^{EF} + {\bf \Omega}^E_{~G} \wedge {\bf \Omega}^{GF},
\end{equation} 
where ${\bf \Omega}^{EF}$, ${\bf T}^E$ and ${\bf R}^{EF}$ are the generalized Levi-Civita connection, torsion and Ricci curvature differential forms in the space-time ${\cal M}^6$.

Since all the formulas are perfect parallelism to ones in Section 2, it is straightforward to calculate the generalized Ricci scalar and define the generalized Hilbert-Einstein action as follows
\begin{eqnarray} \label{6SHE}
	S_{HE}(6) &=& M^2_{Planck}(6) \int d^6x \sqrt{|det ({\bf G})|}~ {\bf  R}_6, \nonumber \\
	{\bf R}_6 &=& \eta^{EG} {\bf R}_{EFGH}(\hat x) \eta^{FH}. 	
\end{eqnarray}

As our purpose is to construct a minimal extension of the Standard Model, we will compute $S_{HE}(6)$ in a specific ansatz of the vielbein in Eq.(\ref{6Vielbein}).


\section{A simplified model}

The general ${\cal M}^6$ vielbein can contain a rich spectrum of fields with many possibilities. Since our goal is to incorporate the Standard Model into this framework, we will restrict the vielbein with an ansatz, which can lead to a minimal extension with reasonable features. 

\subsection{Ansatz and Hilbert-Einstein action}
In this model, we will not identify the vector fields in the ${\cal M}^5$ vielbein $E^{IA}_M(\hat x)$ as the Standard Model's ones. Therefore, the following specialized ansatz is chosen
\begin{eqnarray} \label{6Vielbein1}
		E^{wA}_M (\hat x) &=&  E^{vA}_M (\hat x) = E^{A}_M (\hat x), \nonumber \\
	E^{a}_\mu (\hat x) &=& e^a_\mu(x). {\bf 1}_2,~~ E^a_5(\hat x) = 0 \nonumber \\
	E^{\dot 5}_5 (\hat x) &=& \lambda^2 {\bf 1}_2,\hskip 24pt	E^{\dot 5}_\mu (\hat x) = 0, 
\end{eqnarray}
where $\lambda^2$ is the metric parameter of the chiral dimension.

The components of the inverse vielbein $E^M_A(\hat x)$  in this case is 
\begin{eqnarray} \label{6VielbeinInv}
	E_{a}^\mu (\hat x) &=& e_a^\mu(x). {\bf 1}_2,~~ E^a_5(\hat x) = 0 \nonumber \\
	E_{\dot 5}^5 (\hat x) &=& \lambda^{-2} {\bf 1}_2,\hskip 24pt	E_{\dot 5}^\mu (\hat x) = 0. 
\end{eqnarray}

With the ansatz given in Eq.(\ref{6Vielbein1}) we can calculate the ${\cal M}^6$ metric tensor explicitly for this case as follows 
\begin{eqnarray} \label{6Metric1}
	G^{66}(\tilde x) &=& \begin{bmatrix}
		G^{MN}(\hat x) B_M(\hat x) B_N(\hat x) &0\\
		0& G^{MN}(\hat x) X_M(\hat x) X_N(\hat x) 	
	\end{bmatrix} + \lambda^{-4}_D {\bf 1}_2 \otimes {\bf 1}_2 \nonumber \\
	G^{6M}(\tilde x) &=& \begin{bmatrix}
		G^{M N}(\hat x) B_N(\hat x) & 0 \\
		0 & G^{MN}(x) X_N (\hat x)
	\end{bmatrix},~~G^{MN}(\tilde x)= G^{MN}(\hat x) {\bf 1}_2, \nonumber \\
 G^{\mu \nu}(\tilde x) &=& g^{\mu \nu} (x) {\bf 1}_2 \otimes {\bf 1}_2,~~
	 G^{\mu 5}(\tilde x) = 0,~~ G^{55}(\tilde x) = \lambda^{-4} {\bf 1}_2 \otimes {\bf 1}_2.
\end{eqnarray}

Hence, the $5+1$-dimensional Hilbert-Einstein action is reduced to the $4+1$-dimensional Einstein-Hilbert and Yang-Mills gauge actions as follows 
\begin{eqnarray} 
	S_{HE}(6) &=& M^2_{Planck}(6) \int d^6x \sqrt{|det ({\bf G})|}~ {\bf  R}_6 \nonumber \\
	&=& M^2_{Planck}(6) \lambda^2_D \int d^5x \sqrt{|det(G)|} ~tr {\bf R}_6 = S_{HE}(5) + S_g(5) \label{HE6} \\
	S_{HE}(5) &=& M^2_{Planck}(5) Tr \int d^5x \sqrt{det |G|} R_5  \nonumber \\
	&=& M^2_{Planck}(4) Tr \int d^4x \sqrt{det |g|}~ r_4  \\
	S_g(5) &=&  M^2_{Planck}(4)\int d^4x \sqrt{det |g|}~{\cal L}_g (5) \nonumber \\
	{\cal L}_g (5) &=& - { M^2_{Planck}(4) \over 16} Tr (G^{MK} (\hat x) G^{NL}(x) {\cal F}^\dagger_{MN} (\hat x) {\cal F}_{KL}(\hat x)), 
	\label{Lg5} 
\end{eqnarray}
where ${\bf R_6}$, $R_5$ and $r_4$ are Ricci scalar curvatures, while ${\bf G}$, $G$ and $g$ are metrics respectively in ${\cal M}^6$, $ {\cal M}^5$ and ${\cal M}^4$. The Planck masses are related as follows
\begin{equation} \label{PlanckMassRel}
	M_{Planck}(4) = \lambda M_{Planck}(5)= \lambda \lambda_D  M_{Planck}(6).
\end{equation}

Since the vielbein components are dimensionless, we can introduce the physical vector fields by the scaling redefinitions
\begin{equation}
	B_M(\hat x) \rightarrow {2 \over M_{Planck}(4) f_\kappa} B_M(\hat x),~~ X_M(\hat x) \rightarrow  {2 \over M_{Planck}(4) f_\kappa} X_M(\hat x).  
\end{equation} 
Therefore, the $4+1$-dimensional gauge Lagrangian ${\cal L}_g(5)$ will have the correct factor as follows
\begin{eqnarray} \label{Lg5}
	{\cal L}_g (5) &=& - {1 \over 4 f_\kappa^2 }  Tr (G^{MK} (\hat x) G^{NL}(x)  {\cal F}^\dagger_{MN} (\hat x) {\cal F}_{KL}(\hat x)) \nonumber \\
	&=& ~-{1 \over 4 f^2_\kappa}Tr( g^{\mu \rho}(x) g^{\nu \tau}(x) {\cal F}^\dagger_{\mu \nu} (\hat x) {\cal F}_{\rho \tau}(\hat x) \nonumber \\
	&&~- 2 \lambda^{-4} g^{\mu \nu}(x) {\cal F}^\dagger_{\mu 5} (\hat x) {\cal F}_{\nu 5}(\hat x) + \lambda^{-8} {\cal F}^\dagger_{55} (\hat x) {\cal F}_{55}(\hat x)   ),  
\end{eqnarray}
where $f_\kappa$ is a dimensionless parameter as in Eq.(\ref{Fkappa}). Similarly to Eq.(\ref{CALF}) the "field strength" components ${\cal F}_{MN}(\hat x)$ are given as follows
\begin{eqnarray}
	{\cal F}_{\mu \nu} (\hat x) &=& B_{\mu \nu}(\hat x) + X_{\mu \nu}(\hat x) + ([X_\nu(\hat x), B_\mu(\hat x)] + [B_\nu(\hat x), X_\mu(\hat x)]), \\
	 {\cal F}_{\mu 5} (\hat x) &=& B_{\mu 5}(\hat x) + X_{\mu 5}(\hat x) + ([X_5(\hat x), B_\mu(\hat x)] + [B_5(\hat x), X_\mu(\hat x)]), \\
	 {\cal F}_{55} (\hat x) &=& B_{55}(\hat x) + X_{55}(\hat x) + (\{X_5(\hat x), B_5(\hat x)\} + \{B_5(\hat x), X_5(\hat x)\}),
\end{eqnarray}
where
\begin{eqnarray}
	B_{\mu \nu}(\hat x) &=& \partial_\mu B_\nu(\hat x) - \partial_\nu B_\mu(\hat x) + [B_\mu(\hat x), B_\nu(\hat x)] \\
	X_{\mu \nu}(\hat x) &=& \partial_\mu X_\nu(\hat x) - \partial_\nu X_\mu(\hat x) + [X_\mu(\hat x), X_\nu(\hat x)]
\end{eqnarray}

We have also introduced the dimensionless parameter $f_\kappa$ as follows
\begin{equation}
	 f_\kappa = {4m \over M_{Planck}(4)}. \label{mMPlanck}
\end{equation}

Similarly to what has been shown in \cite{VDHW2017, VietDu2017}, the field strength ${\cal F}_{\mu \nu}$ is gauge covariant if 
\begin{equation} \label{BXCom}
	[B_\mu(\hat x), X_\nu(\hat x)] = [X_5(\hat x), B_\mu(\hat x))] = [X_\mu(\hat x), B_5(\hat x)] = 0.
\end{equation} 
That is to say $X_\mu(\hat x)$ is an abelian gauge field.

\subsection{The Standard Model's gauge sector and predictions}

The gauge Lagrangian ${\cal L}_g (5)$ in Eq.(\ref{Lg5}) can be split into the following three terms 
\begin{equation}
	{\cal L}_g (5)
	= {\cal L}_g(B) + {\cal L}_g (X) + {\cal L}_g(B,X),
\end{equation}
where ${\cal L}_g(B)$ and ${\cal L}_g (X)$ contain respectively only the ${\cal M}^5$-vectors $B_M(\hat x)$ and $X_M(\hat x)$, while ${\cal L}_g(B,X)$ is the cross interactions between these two fields.

In this model, the Standard Model's gauge vector and Higgs fields, including the gluon ones, now can be assigned to the components of $B_M(\hat x)$ as in Eqs.(\ref{SMHiggs}) and (\ref{StrongGauge}) in Section 2.

Therefore, one can repeat the calculation in Section 2, the Lagrangian ${\cal L}_g(B)$ modifies the one in Eq.(\ref{Fkappa}) with the parameter $\lambda$ as follows 
\begin{eqnarray}
&{\cal L}_g(B) &= - {1 \over 4f^2_\kappa} ( g^{\mu \rho}(x) g^{\nu \tau}(x) tr (B^\dagger_{\mu \nu}(\hat x)  B_{\rho \tau} (\hat x)) \nonumber \\ 
&& - 2 {\lambda^{-4}} g^{\mu \nu}(x) tr( {\cal B}^\dagger_{5\mu}(\hat x) {\cal B}_{5 \nu}(\hat x)) +  {\lambda^{-8}}  tr({\cal B}^\dagger_{55}(\hat x) {\cal B}_{55}(\hat x)).	
\end{eqnarray}

So, the Standard Model's gauge and Higgs Lagrangians are recovered with the following restrictions
\begin{eqnarray}
	f_\kappa &=& {4 m \over M_{Planck}(4)} = 2 g \sqrt{N_F} \lambda \label{mMPlanck2} \\
	\lambda^4 N_F &=&  Tr (\bf Y^\dagger Y) \\
	3g_S/2&=& g =~ g' \sqrt{10 \over 3}. \label{gprime}
\end{eqnarray}

In this model, all the Standard Model's coupling constants are characterized by a single one $g$ due to Eq.(\ref{gprime}). 

Eq.(\ref{gprime}) leads to the same  Weinberg angle of $\sin^2 \theta_W=0.23077$ as in Section 2. The top quark and Higgs masses are predicted as follows 
\begin{eqnarray}
	m_t &=& 2 m_W \lambda = \sqrt{2} m_1 \lambda  \\
	m_H &\leq& m_W \sqrt{N_F} = 139~GeV.
\end{eqnarray}
Hence, the top quark mass can be fitted with its experimental value of $173~MeV$, when $\lambda=1.075$ is chosen. The upper bound of Higgs mass remains unchanged as $139~MeV$. The mass parameter $m_1$ can also be determined as $113~ MeV$.

The mass parameter $m$ now is related to the Planck mass due to Eq.(\ref{mMPlanck2})
\begin{equation}
	m = {g \sqrt{N_F} \lambda \over 2} M_{Planck}(4) = {\sqrt{4 \pi \alpha N_F} \lambda \over 2 \sin \theta_W} M_{Planck} = 0.587 M_{Planck}(4).
\end{equation}

\subsection{The dark gauge sector}
The most general abelian ${\cal M}^5$ dark gauge vector $X_M(\hat x)$, satisfying Eq.(\ref{BXCom}) is given as follows
\begin{eqnarray} \label{Xgauge}
	X_\mu(\hat x) &=&  \begin{bmatrix}
		g_V~ V_\mu(x) {\bf 1}_{8N_F} & 0 \\
		0 &  g'_V~ V_\mu(x) {\bf 1}_{8N_F} + g_D D_{\mu}(x) {\bf Q}_{D}
	\end{bmatrix}, \nonumber \\
    X_5(\hat X) &=& \begin{bmatrix}
    	X_5(x) & 0 \\
    	0 & X^\dagger_5(x)
    \end{bmatrix} \\
	X_5 (\hat x) &=& g  \begin{bmatrix}
		\phi(x) - {m_2 \over g} & \\
		0 & \phi(x) -{m_2 \over g}
	\end{bmatrix} \otimes f_\phi {\bf 1}_{8N_F} ,
\end{eqnarray}
where $V_\mu(x)$ and $D_\mu(x)$ are vector fields. The dark charge operator ${\bf Q}_D$ must be diagonal. $f_\phi$ is the Yukawa coupling constant of the scalar Higgs field $\phi(x)$.

It is straightforward to calculate the gauge Lagrangian of the dark vector field $X_M(\hat x)$ and show that $V_\mu(x)$ is massless. If we identify $V_\mu(x)$ with the electromagnetic field, it will alter the good prediction of the Weinberg angle. Therefore, for simplicity, we consider $X_\mu(\hat x)$ in the following form
\begin{eqnarray}
	X_\mu(\hat x) &=&  \begin{bmatrix}
		0 & 0 \\
		0 &  g_D D_{\mu}(x) {\bf Q}_{D}
	\end{bmatrix}. 
\end{eqnarray}

The gauge Lagrangian of the dark sector derived from the generalized Hilbert-Einstein Lagrangian ${\cal L}_{HE}(6)$ is 
\begin{eqnarray}	
	&{\cal L}_g(X) &= - {1 \over  f^2_\kappa} (g^{\mu \rho}(x) g^{\nu \tau}(x) tr( {\cal D}^\dagger_{\mu \nu}(\hat x)  {\cal D}_{\rho \tau} (\hat x)) \nonumber   \\
	&& - 2 {\lambda^{-4}} g^{\mu \nu}(x) tr( {\cal D}^\dagger_{5\mu}(\hat x) {\cal D}_{5 \nu}(\hat x)) +  {\lambda^{-8}}  tr({\cal D}^\dagger_{55}(\hat x) {\cal D}_{55}(\hat x)),   
\end{eqnarray}  
where the field strength of $X_M(\hat x)$ is given as follows
\begin{eqnarray}
	{\cal X}_{\mu \nu} (\hat x) &=&- {g_D \over 2}(\partial_\mu D_{\nu}(x) - \partial_\nu D_\mu(x)) \otimes \begin{bmatrix}
		0 & 0 \\
		0 & 1
	\end{bmatrix} \otimes Q_D, \nonumber \\
	{\cal X}_{\mu 5} (\hat x) &=& {g f_\phi \over 2}(\partial_\mu. {\bf 1}_{16N_F} +
	g_D D_\mu (x) {\bf Q}_D \otimes \sigma_3)\phi(x), \nonumber \\
	{\cal X}_{55} (\hat x) &=& g^2 (\phi^2(x) - {m^2_2 \over g^2} ) f^2_\phi \otimes {\bf 1}_{8N_F}.\label{Xstrength}
\end{eqnarray}

It is straightforward to calculate the Lagrangian ${\cal L}_g(X)$ as follows
\begin{eqnarray}\label{LgX1}
	{\cal L}_g(X) &=&  -~ {1 \over 4} g^{\mu \rho} (x) g^{\nu \tau}(x) D_{\mu \nu}(x) D_{\rho \tau}(x) + {1 \over 2} g^{\mu \nu}(x)\partial_\mu \phi(x) \partial_\nu \phi(x) \nonumber \\
	&& + {1 \over 2} g^2_D g^{\mu \nu}(x) D_\mu(x) D_\nu(x) \phi^2(x) - {g^4} N_F(\phi^2(x) - {m^2_2 \over g^2})^2, 
\end{eqnarray}
where $D_{\mu \nu}(x) = \partial_\mu D_\nu(x) - \partial_\nu D_\mu(x)$.

In Eq.(\ref{LgX1}), we have imposed the following conditions to have correct factors for the kinetic terms
\begin{eqnarray} 
	& f_\phi& =~  {\lambda^2 \over 2} =0.5778, \label{fphi}\\
	&4N_F g^2&=~ g^2_D Tr(Q^2_D) \label{gD}
\end{eqnarray}

The vector field $D_\mu(x)$ will acquire a mass $m_D = g_D m_2/g$ due to the vacuum expectation value of $\phi(x)$ $v_\phi= m_2/g$. So, the massive vector $V_\mu(x)$ would be a good candidate for the so-called dark photon. 

\subsection{Mixing of the dark photon with the Standard Model gauge sector}

The mixed interaction of the generalized vectors fields $B_M(\hat x)$ and $X_M(\hat x)$ consists of the following three terms  
\begin{eqnarray}
		&{\cal L}_g(B,X)&= {\cal L}_g(B,X) (1) + {\cal L}_g(B,X) (2)+ 	{\cal L}_g(B,X) (3) \nonumber \\
	&{\cal L}_g(B,X) (1) &= - {1 \over f^2_\kappa} g^{\mu \rho}(x) g^{\nu \tau}(x) tr(B^\dagger_{R\mu \nu}(x)  {\cal D}_{\rho \tau} (\hat x) + {\cal D}^\dagger_{\mu \nu}(x)  B_{R\rho \tau} (x)) \nonumber \\
	&{\cal L}_g(B,X) (2)
	&=  {2\lambda^{-4} \over f^2_\kappa} g^{\mu \nu}(x) Tr({\cal B}^\dagger_{R5\mu}(x) {\cal D}_{5\nu} (x) + {\cal D}^\dagger_{5\mu}(x) {\cal B}_{R5 \nu} (x) )  \nonumber \\
	&{\cal L}_g(B,X) (3)	&= - {\lambda^{-8} \over f^2_\kappa} Tr({\cal X}^\dagger_{55}(\hat x) {\cal B}_{55} + {\cal B}^\dagger_{55}(\hat x) {\cal X}_{55} (\hat x) ).
\end{eqnarray}

Since only the right-handed gauge fields of the Standard Model will contribute to the mixed Lagrangian ${\cal L}_g(B,X)$, we have calculated it explicitly as follows
\begin{eqnarray}
	{\cal L}_g(B,X) &=& \epsilon g^{\mu \rho} (x) g^{\nu \tau}(x) W^0_{\mu \nu}(x) X_{\rho \tau}(x) + \epsilon' g^{\mu \nu}(x) Tr (\nabla_\mu \phi(x) \nabla_\nu {\hat H}(x))  \nonumber \\
	&& -4g^2 N_F (\phi^2 - {m^2_2 \over g})(\bar H(x) H(x) - {2 m^2_1 \over g^2}),	
\end{eqnarray}
where
\begin{eqnarray}
\epsilon &=& - {g' g_D Tr(Y_R Q_D) \over 8 N_F g^2}, \label{epsilon}\\
\epsilon'&=& 4\epsilon {g^2 \over \lambda^4 g' g_D}.	
\end{eqnarray}

The kinetic gauge and Higgs mixing parameters $\epsilon$ and $\epsilon'$ depend on the unknown dark charge operator $Q_D$. It is commonly believed that $\epsilon < 10^{-3}$. 
The simplest way to satisfy the upper bound is to assume that only the right-handed neutrinos and their Kaluza-Klein can have a dark charge. In that case, we have the following restrictions  
\begin{eqnarray}
\epsilon &=& \epsilon' = 0,\\
g_D &=& 2g.
\end{eqnarray}
The mixed Lagrangian is simplified to
\begin{eqnarray}
	{\cal L}_g(B,X) &=&  -4g^2 N_F (\phi^2(x) - {m^2_2 \over g^2})(\bar H(x) H(x) - {2 m^2_1 \over g^2}).	
\end{eqnarray}

In summary, we have derived the Standard Model and Einstein's gravity together with a dark photon and a dark Higgs fields as components of the generalized gravity in ${\cal M}^6$.

\subsection{Quark-lepton mixing with the dark sector and the mass parameters}

In our framework, it is possible to derive the couplings of gauge and Higgs fields to fermions can from the Lagrangian
\begin{equation}
	{\cal L}_{f-g} = iTr  {\bar {\bf \Psi}}(\tilde x) \slashed {\bf D} {\bf \Psi} (\tilde x) = iTr  {\bar {\bf \Psi}}(\tilde x)  {\bf \Gamma}^E {\bf E}^P_E(\tilde x) D_P {\bf \Psi}(\tilde x) , 
\end{equation}
which is generalization of ${\cal L}_D$ in Eq.(\ref{DiracL}). 
Let us expand the curved Dirac operator's terms in the dark space into the kinetic, mass, and gauge coupling terms as follows
\begin{eqnarray}
	i\slashed {\bf D} &=& D_0 + D_{m}  + D_{g}  \nonumber \\
	D_0 &=& i{\bf \Gamma}^a {\bf E}^\mu_a (\tilde x) {\bf D}_\mu =i \gamma^a e^\mu_a(x) \partial_\mu  \otimes {\bf 1}_2 \otimes {\bf 1}_2 = \slashed \partial \otimes {\bf 1}_2 \otimes {\bf 1}_2 \nonumber \\
	D_m &=& {\bf 1}_4 \otimes \sigma_2 \otimes {\cal M}\\
	D_{g}  &=& i{\bf \Gamma}^A {\bf E}_A^6 (\tilde x) {\bf D}_6  =  {\bf \Gamma}^A  E^{M}_A(\hat x) \begin{bmatrix}
		B_M(\hat x) & 0 \\
		0 &   X_M(\hat x)
	\end{bmatrix},
\end{eqnarray}
where the mass matrix ${\cal M}$ replaces the flat one in Eq.(\ref{M0M}) with the metric parameters $\lambda$ and $\lambda_D$ as follows
\begin{equation}
	{\cal M} = \begin{bmatrix}
		m_1/\lambda^2 & - i m /\lambda_D^2 \\
		i m/\lambda_D^2 & m_2 /\lambda^2
	\end{bmatrix}.
\end{equation}

$D_0$ is the usual Dirac operator of quark-leptons and their dark partners, which includes gravity via the usual vierbein $e^\mu_a(x)$. The couplings of gauge and Higgs fields to fermions in both dark and Standard Model's sectors are given in $D_g$.

Like in the flat case, the non-diagonal mass matrix ${\cal M}$ can be diagonalized by the unitary transformation (\ref{Utrans}) leading to the new mass splitting formula
\begin{equation}
	\mu_D-\mu = {2 m \over \lambda^2_D \sin 2\theta}.  
\end{equation}

The mass eigenvalues $\mu$ and $\mu_D$ can be expressed in terms of $m, m_1$ and $\theta$ as follows
\begin{equation}
	\mu= {m_1 \over \lambda^2} - {m \over \lambda^2_D} \tan \theta,~~ \mu_D = {m_1 \over \lambda^2} + {m \over \lambda^2_D} \cot \theta.
\end{equation}

Since in the Standard Model quark-lepton masses are completely generated by the Yukawa couplings with the Higgs field, we will assume $\mu =0$. Therefore,
we have the following relations
\begin{eqnarray}
	\mu_D &=&  {1.174 M_{Planck}(4) \over \lambda^2_D \sin 2\theta}, ~~
	m_2 = m_1 \cot^2 \theta, \nonumber \\
	m_1 &=& \mu_D \lambda^2 \sin^2 \theta= 1.1556 \mu_D \sin^2  \theta, \label{massparameters}
\end{eqnarray}  

The mass parameters $m_2$ and $\mu_D$ determine the dark matter energy scale, depending on the metric parameter $\lambda^2_D$ and the mixing angle $\theta$, while $m_1 = 113~GeV$ determines the weak scale.

\subsection{Gauge mixing between the Standard Model and the dark sectors}

The operator $D_0$ is unchanged during the mixing transformation $U$, giving the usual Dirac kinetic terms for quark-leptons and their dark partners.

The operator $D_{g}$ will transform due to the mixing transformation in Eq.(\ref{Utrans}) as follows
\begin{eqnarray}
	&U D_{g} U^\dagger =  \begin{bmatrix}
		\cos^2 \theta \slashed B(\hat x) + \sin^2 \theta \slashed X(\hat x) & {i \over 2} \sin 2 \theta (\slashed B (\hat x)+ \slashed  X (\hat x)) \\
		- {i \over 2} \sin 2 \theta (\slashed B(\hat x) + \slashed X(\hat x)) & 	 \sin^2 \theta \slashed  B( \hat x) + \cos^2 \theta \slashed X (\hat x)  	 
	\end{bmatrix}.~~& \label{gMix2}
\end{eqnarray}

All the gauge interactions of quark-leptons and their Kaluza-Klein partners are given in the Lagrangian
\begin{equation} \label{Lgf}
	{\cal L}_{f-g} = Tr \bar {\bf \Psi}(\tilde x) D_g {\bf \Psi}(\tilde x).
\end{equation}

First of all, it is straightforward to show that all the interaction terms of fermions with the scalar Higgs field $\phi(x)$ derived from Eq.(\ref{Lgf}) will vanish since they always come out in the form $\bar \psi_{L,R}(x) \phi(x) \psi_{L,R}(x)$. So, the scalar Higgs field does not generate masses to the fermions.


Therefore, in the quark-lepton sector, all the gauge interactions are given by the following Lagrangian
\begin{eqnarray}
	{\cal L}_g(\Psi) &=& \cos^2 \theta~ Tr {\bar \Psi}(\hat x) \slashed B(\hat x) \Psi(\hat x)+ \sin^2 \theta ~Tr {\bar \Psi}(\hat x) \slashed X(\hat x)\Psi(\hat x).~~  
\end{eqnarray}
The first term has been shown in Section 2, to be the Standard Model interactions of quark-leptons with the gauge and Higgs fields with an overall factor of $\cos^2 \theta \sim 1$. In our model, the direct interactions between the dark photon and Higgs field with quark-leptons are allowed with a small factor $2g \sin^2 \theta$. Since these terms replace the kinetic gauge mixing one, the dark photon search's observations at NA64 project \cite{NA64-2019} will be intact if we choose the mixing angle $\theta< 0.032, \sin^2 \theta < 10^{-3}$. The dark photon mass now is $2m_2 \sim 2 m_1/\sin^2 \theta > 226~ TeV$ as a consequence of Eq.(\ref{massparameters}). With the metric parameter $\lambda_D > 10^9$, the Stueckelberg mass of the dark quark-lepton can be in the range $98~TeV < \mu_D < 1000~TeV$.

The gauge interactions of dark quark-leptons are given by the Lagrangian 
\begin{eqnarray}
	{\cal L}_g(\Psi_D) &=& \sin^2 \theta~ Tr {\bar \Psi}_D(\hat x) \slashed B(\hat x) \Psi_D(\hat x)+ \cos^2 \theta ~Tr {\bar \Psi}_D(\hat x) \slashed X(\hat x)\Psi_D(\hat x).~~  
\end{eqnarray}
That is to say, the interactions of quark-leptons'Kaluza-Klein partners with the Standard Model's gauge and Higgs fields will be reduced by the factor $\sin^2 \theta  < 10^{-3}$. This property suggests that the components of $\Psi_D(\hat x)$ behave like dark fermions. The interaction of dark quark-leptons with the dark photon will be $g_D = 2 g$ if only neutrinos have a dark charge as we have assumed above. 

The gauge vector and Higgs fields of both the Standard Model and dark sector can be portal between the quark-lepton and dark fermion sectors as we have the Lagrangian
\begin{eqnarray}
	{\cal L}_g(\Psi,\Psi_D) &=&  {i \over 2} \sin 2\theta ~ Tr {\bar \Psi} (\hat x) (\slashed B(\hat x) +  \slashed X(\hat x))\Psi_D(\hat x) + h.c.~~  
\end{eqnarray}
Due to the factor $\sin 2 \theta < 0.064$, the coupling of one certain quark-lepton and its Kaluza-Klein partner with the Standard Model's gauge vector fields is also small. This term predicts a decay channel of a dark quark-lepton into a vector boson and a quark-lepton with a small rate.

\section{Summary and discussions}
In this paper, we have constructed the Einstein-Dirac system in the space-time ${\cal M}^6$, which is an extension of the usual Riemannian manifold ${\cal M}^4$ with two extra discrete dimensions, each having two points. In addition to the chiral dimension proposed originally by Connes and Lott \cite{CoLo1989}, we have introduced the dark one. The model is based on a single principle of General Relativity extended to space-time with discrete extra dimensions. However, it can lead to a rich structure, where different ideas can emerge naturally.

The most remarkable result is that all the gauge vector and Higgs fields of the Standard Model can be derived as components of the generalized vielbein of the extended space-time without unwanted infinite towers of massive modes. So all the gauge vector and Higgs fields are Kaluza-Klein partners of gravity in some sense. The discreteness of the extra dimension allow them to have different gauge structures.

In this framework, all coupling constants are related to the electroweak one $g$. Thus, the Weinberg angle is predicted with high precision. The top quark mass can be fitted to the experimental value of $173~GeV$ with a choice of the chiral metric parameter $\lambda=1.075$. The Higgs mass is restricted by an upper bound of $139~GeV$ in accordance with the experimental value. The gauge coupling constant $g$ of the Standard Model is related to the Planck mass $M_{Planck}$ and the scale parameter $m$ of the dark dimension by $g \sim 1.074 m/ M_{Planck}$. As Kaluza-Klein partners of the Standard Model's bosons in the dark dimension, in a minimal extension, the extended vielbein also contains a  vector field together with a scalar Higgs field, which has a quartic potential. The scalar Higgs field generates a mass to the dark vector field, but not to any other fields. 

As a consequence of the dark dimension, due to a non-diagonal mass matrix of fermions, quark-leptons will mix with their Kaluza-Klein partners, leading to a weak coupling of quark-leptons with the dark massive vector boson, depending on a small mixing angle. That's why it can be a candidate for a new dark photon, which has some new features other than the usual one. With a simplified assumption that the dark photon interacts only with neutrinos, we can show that all the kinetic gauge mixing terms vanish. Inversely, the coupling of the Standard Model with the Kaluza-Klein partners of quark-leptons is also weak. So, these particles can be called dark quark-leptons. The idea of considering the lightest Kaluza-Klein particles as the dark matter has been investigated \cite{ChengFeng2002, Flacke2017} with the continuous UED dimension \cite{Appequist2000}.

Our framework has three mass parameters. The first one $m$ is related to the Planck mass. The second one $m_1$, which characterized the electroweak scale, is determined as $113~ MeV$. The third one $m_2$, which characterizes the dark scale, depends on the mixing angle. With a reasonable choice of it, $m_2$ will be at least a few hundred $TeV$.  

The last parameter in our framework is the dark metric one $\lambda_D$, which must be at least $10^{9}$ to have the dark energy scale at the $TeV$ range. It is compatible with the Randall-Sundrum model warp factor $\sim 45$ \cite{RaSu1999, RaSu1999a}.

There are still questions in our model to be addressed in future research.
The first one is the relation between strong and electroweak coupling constants $g_S= 2g/3$. This restriction is also common to all the noncommutative geometric Standard Models \cite{Besnard}. Since the predicted value of the Weinberg angle suggests that the model's energy scale is the electroweak one, this restriction would require a fix. It means that the incorporation of the color and electroweak fields in the gauge sector must be more sophisticated than in Eq.(\ref{StrongGauge}). It is likely that more discrete extra dimensions could be introduced. 

The second one is the decay channel of dark quark-lepton into a photon and a quark-lepton. If one wants to bring the dark scale down to the $TeV$ range by choosing the mixing angle $\sin^2 \theta_W \sim 10^{-3}$ the phenomenological consequences of this decay must be investigated carefully./.  

The research is funded by Vietnam National Foundation for Science and Technology Development(NAFOSTED) under grant No 103.01-2017.319.

\bibliography{6DReferences}

\end{document}